\documentclass[aps,prd,showpacs,superscriptaddress,nofootinbib,twocolumn,10pt]{revtex4-2}
\usepackage{color,amsmath,amssymb,graphicx,latexsym,lineno}
\usepackage{threeparttable,multirow,txfonts,subfigure,booktabs}
\graphicspath{{./}{figure_for_paper/}}

\begin{document}

\title{Measurement of ultra-high-energy diffuse gamma-ray emission of the Galactic plane 
from 10 TeV to 1 PeV with LHAASO-KM2A}

\author{Zhen Cao}
\affiliation{Key Laboratory of Particle Astrophyics \& Experimental Physics Division \& Computing Center, Institute of High Energy Physics, Chinese Academy of Sciences, 100049 Beijing, China}
\affiliation{University of Chinese Academy of Sciences, 100049 Beijing, China}
\affiliation{Tianfu Cosmic Ray Research Center, 610000 Chengdu, Sichuan,  China}

\author{F. Aharonian}
\affiliation{Dublin Institute for Advanced Studies, 31 Fitzwilliam Place, 2 Dublin, Ireland }
\affiliation{Max-Planck-Institut for Nuclear Physics, P.O. Box 103980, 69029  Heidelberg, Germany}

\author{Q. An}
\affiliation{State Key Laboratory of Particle Detection and Electronics, China}
\affiliation{University of Science and Technology of China, 230026 Hefei, Anhui, China}

\author{Axikegu}
\affiliation{School of Physical Science and Technology \&  School of Information Science and Technology, Southwest Jiaotong University, 610031 Chengdu, Sichuan, China}

\author{Y.X. Bai}
\affiliation{Key Laboratory of Particle Astrophyics \& Experimental Physics Division \& Computing Center, Institute of High Energy Physics, Chinese Academy of Sciences, 100049 Beijing, China}
\affiliation{Tianfu Cosmic Ray Research Center, 610000 Chengdu, Sichuan,  China}

\author{Y.W. Bao}
\affiliation{School of Astronomy and Space Science, Nanjing University, 210023 Nanjing, Jiangsu, China}

\author{D. Bastieri}
\affiliation{Center for Astrophysics, Guangzhou University, 510006 Guangzhou, Guangdong, China}

\author{X.J. Bi}
\affiliation{Key Laboratory of Particle Astrophyics \& Experimental Physics Division \& Computing Center, Institute of High Energy Physics, Chinese Academy of Sciences, 100049 Beijing, China}
\affiliation{University of Chinese Academy of Sciences, 100049 Beijing, China}
\affiliation{Tianfu Cosmic Ray Research Center, 610000 Chengdu, Sichuan,  China}

\author{Y.J. Bi}
\affiliation{Key Laboratory of Particle Astrophyics \& Experimental Physics Division \& Computing Center, Institute of High Energy Physics, Chinese Academy of Sciences, 100049 Beijing, China}
\affiliation{Tianfu Cosmic Ray Research Center, 610000 Chengdu, Sichuan,  China}

\author{J.T. Cai}
\affiliation{Center for Astrophysics, Guangzhou University, 510006 Guangzhou, Guangdong, China}

% ...
\collaboration{The LHAASO Collaboration}

\email[E-mail: ]{xyhuang@pmo.ac.cn; xisq@ihep.ac.cn; yuyh@ustc.edu.cn; yuanq@pmo.ac.cn; zhangrui@pmo.ac.cn; zhangyi@pmo.ac.cn; zhaosp@mail.sdu.edu.cn}

\date{\today}

\begin{abstract}
The diffuse Galactic $\gamma$-ray emission, mainly produced via interactions 
between cosmic rays and the interstellar medium and/or radiation field, 
is a very important probe of the distribution, propagation, and interaction of 
cosmic rays in the Milky Way. In this work we report the measurements of diffuse 
$\gamma$-rays from the Galactic plane between 10 TeV and 1 PeV energies, with 
the square kilometer array of the Large High Altitude Air Shower Observatory 
(LHAASO). Diffuse emissions from the inner ($15^{\circ}<l<125^{\circ}$, 
$|b|<5^{\circ}$) and outer ($125^{\circ}<l<235^{\circ}$, $|b|<5^{\circ}$) 
Galactic plane are detected with $29.1\sigma$ and $12.7\sigma$ significance, 
respectively. The outer Galactic plane diffuse emission is detected for the 
first time in the very- to ultra-high-energy domain ($E>10$~TeV). The energy 
spectrum in the inner Galaxy regions can be described by a power-law function 
with an index of $-2.99\pm0.04$, which is different from the curved spectrum as 
expected from hadronic interactions between locally measured cosmic rays and the 
line-of-sight integrated gas content. Furthermore, the measured flux is higher 
by a factor of $\sim3$ than the prediction. A similar spectrum with an index of 
$-2.99\pm0.07$ is found in the outer Galaxy region, and the absolute flux for 
$10\lesssim E\lesssim60$ TeV is again higher than the prediction for hadronic 
cosmic ray interactions. The latitude distributions of the diffuse emission are 
consistent with the gas distribution, while the longitude distributions show 
clear deviation from the gas distribution. The LHAASO measurements imply 
that either additional emission sources exist or cosmic ray intensities have 
spatial variations.
\end{abstract}

\pacs{95.85.Pw,98.70.Sa}

\maketitle
The origin and propagation of cosmic rays (CRs) remain among the most
important unresolved problems in astrophysics. Unlike the direct measurements
of energy spectra and anisotropies of CRs in the local vicinity, the diffuse
Galactic $\gamma$-ray emission allows a measurement of the spatial
distribution of CRs throughout the Galaxy. It can thus provide much more
important information of the production and propagation of CRs.
Typically there are three main components of the diffuse Galactic emission
\cite{2000A&A...362..937A,2000ApJ...537..763S,2004ApJ...613..962S}: the decay 
of neutral pions produced by inelastic collisions between CR nuclei and the 
interstellar medium (ISM), the inverse Compton scattering (ICS) of CR $e^{\pm}$ 
off the interstellar radiation field (ISRF), and the bremsstrahlung radiation 
of $e^{\pm}$ in the ISM. The canonical CR propagation and interaction model 
(homogeneous and isotropic diffusion) can largely account for the all-sky 
data measured by space telescopes while being consistent with the local CR 
measurements, except for the underpredicted $\gamma$-ray fluxes in the inner 
Galaxy for energies above a few GeV
\cite{2000ApJ...537..763S,2004ApJ...613..962S,2012ApJ...750....3A}.

For energies below 1 TeV, the all-sky diffuse emission has been measured by
space detectors such as OSO-3 \cite{1968ApJ...153L.203C},
SAS-2 \cite{1975ApJ...198..163F}, COS-B \cite{1982A&A...105..164M}, 
EGRET \cite{1997ApJ...481..205H}, and Fermi-LAT \cite{2012ApJ...750....3A}. 
At higher energies, successful detections of the diffuse emission were only 
achieved by a few ground-based experiments in selected regions of the Galactic plane
\cite{2007ApJ...658L..33A,2008ApJ...688.1078A,2014PhRvD..90l2007A,
2015ApJ...806...20B,HAWC:2021bvb,2021PhRvL.126n1101A}.
The recent measurements of the diffuse emission above 100 TeV by Tibet-AS$\gamma$
\cite{2021PhRvL.126n1101A} revealed flux excesses compared with the conventional model 
prediction (e.g., \cite{2021ApJ...914L...7L,2021PhRvD.104d3010K,2022FrPhy..1744501Q}).
High-precision measurements of the diffuse emission in the very-high-energy
(VHE; 30 GeV to 30 TeV \cite{2004vhec.book.....A}) to ultra-high-energy 
(UHE; 30 TeV to 30 PeV \cite{2004vhec.book.....A}) domain, 
with only minor statistical and systematic uncertainties, are crucial to 
understanding the origin and propagation of CRs, particularly the physical 
origin of the new spectral features of CR nuclei by recent direct measurements
\cite{2011Sci...332...69A,2015PhRvL.114q1103A,2019SciA....5.3793A,2021PhRvL.126t1102A}
and the potential contributions from unresolved source populations (e.g., \cite{2000A&A...362..937A,2004vhec.book.....A,2018PhRvL.120l1101L, 2020A&A...643A.137S,2022ApJ...928...19V,2022PhRvD.105b3002Z}).

We report the measurements of the diffuse emission from the Galactic plane
in a wide energy range, from 10 TeV to 1000 TeV. We use the data recorded by
the square kilometer array (KM2A) of the Large High Altitude Air Shower
Observatory (LHAASO) experiment located at Haizi Mountain ($100^{\circ}\!.01$E,
$29^{\circ}$\!.35N; 4400 m above the sea level), Daocheng, Sichuan province,
China \cite{2019arXiv190502773C}. The LHAASO experiment is a large area, 
wide field-of-view observatory for CRs and $\gamma$-rays with hybrid detection 
techniques \cite{2019arXiv190502773C}. In April 2019, LHAASO started taking 
data with a partial array, and successfully opened the PeV $\gamma$-ray 
window with its extraordinary sensitivity
\cite{LHAASO_nature,LHAASO_science,2021PhRvL.126x1103A,2021ApJ...919L..22C,2021ApJ...917L...4C}.
In July 2021, LHAASO completed the installation of its entire detector array 
and started its scientific operation.

%It consists of three sub-detector arrays: 1) the KM2A, which is made up of
%5195 electromagnetic detectors (EDs) and 1188 muon detectors (MDs), covers an 
%area of $\sim1.3$ km$^2$ \cite{2021ChPhC..45b5002A}; 2) the water Cherenkov 
%detector array (WCDA) with three water pools covering 78,000 m$^2$ detection
%area \cite{2021ChPhC..45h5002A};3) The wide field-of-view Cherenkov telescope
%array (WFCTA) with 18 telescopes \cite{2021EPJC...81..657A}. As the most 
%sensitive detector for UHE $\gamma$-rays and CRs, LHAASO is expected to give 
%revolutionary insights into astroparticle physics, significantly advancing 
%the studies of CR physics, $\gamma$-ray astronomy, and new physics.

Air showers produced by primary particles were simulated with the CORSIKA
code (version 7.6400) \cite{1998Heck}, with the GHEISHA and QGSII models for 
low-energy ($\leq80$ GeV) and high-energy ($>80$ GeV) hadronic interactions. 
We simulate $\gamma$-ray events, with energies from 1 TeV to 10 PeV following 
a power-law spectrum of $E^{-2.0}$, with zenith angles from $0^{\circ}$ to
$70^{\circ}$, and with shower cores randomly distributed within 1 km from
the array center. The detector response of KM2A was simulated using a
speciﬁc software G4KM2A \cite{2019ICRC...36..219C}, which is based on the
Geant4 framework (v4.10.00) \cite{2003Agostinelli}. The simulation data
were reconstructed using the same algorithms applied to the experimental
data. Comparisons between simulation and observational data show good
consistency \cite{2021ChPhC..45b5002A}.

{\it Data analysis.} ---
The present work utilizes data acquired by the KM2A 1/2 array from December 26, 
2019 to November 30, 2020 (with a live time of 302 days), by the 3/4 array from 
December 1, 2020 to July 19, 2021 (with a live time of 219 days), and by the full 
array from July 20, 2021 to September 30, 2022 (with a live time of 423 days). 
The whole Galactic plane with Galactic latitudes within $\pm5^{\circ}$ in the 
field-of-view of LHAASO is adopted as our region of interest (ROI). We further
divide the analysis region into two parts, the inner Galaxy region 
($|b| < 5^{\circ}$, $15^{\circ}<l<125^{\circ}$) and outer Galaxy region 
($|b|< 5^{\circ}$, $125^{\circ}<l<235^{\circ}$), for detailed studies.

%The data used in this work were collected by the 1/2 array of KM2A from
%December 26, 2019 to November 30, 2020 with a live time of about 302 days,
%by the 3/4 array from December 1, 2020 to July 19, 2021 with a live time 
%of 219 days, and by the full array from July 20 2021, to September 30, 2022 
%with a live time of 423 days. The regions of interest (ROI) of this work 
%include two regions in the Galactic plane, the inner Galaxy region
%($|b|\leq5^{\circ}$, $15^{\circ}<l<125^{\circ}$) and the outer Galaxy
%region ($|b|\leq5^{\circ}$, $125^{\circ}<l<235^{\circ}$).

We use the sequence of arrival times and deposited energy of the secondary 
particles recorded by the electromagnetic detectors (EDs) to reconstruct 
the direction and energy of a primary CR or $\gamma$-ray event. 
The 68$\%$ containment angle of KM2A is about $0.5^{\circ}-0.8^{\circ}$ 
at 20 TeV and $0.24^{\circ}-0.30^{\circ}$ at 100 TeV, depending on the 
zenith angles of incident photons \cite{2021ChPhC..45b5002A}. The energy 
is reconstructed using the particle density at a perpendicular distance 
of 50 m from the shower axis, $\rho_{50}$, as obtained via a likelihood 
fit of the lateral distribution  using the Nishimura-Kamata-Greisen
(NKG; \cite{1960Greisen}) function. The energy resolution is about 24$\%$ 
at 20 TeV and 13$\%$ at 100 TeV, for events with zenith angles less than
20$^{\circ}$ \cite{2021ChPhC..45b5002A}.

The numbers of muons recorded by the muon detectors (MDs) were used to 
distinguish $\gamma$-ray-induced showers from CR-induced ones. 
The muon-to-electromagnetic-particle ratio, $R_{\mu e}=\log[(N_{\mu}+10^{-4})/N_e]$, 
was used to reject CR events. For point-like source analysis, the muon selection 
criteria were optimized to guarantee retaining 
90\% of the $\gamma$-ray events at energies above 100 TeV for a typical 
Crab-like source based on simulations \cite{2021ChPhC..45b5002A}. In this work, 
we restricted criteria to further suppress the CR background at high energies, 
thus obtaining a maximal detection significance of the diffuse emission. The CR 
rejection power is about $98\%$ at 10 TeV and $99.9993\%$ at 400 TeV, and the 
$\gamma$-ray survival fraction is higher than $50\%$ for the interest energy 
range in this work.

We further apply the following event selection conditions: 1) the number of 
triggered EDs and the number of deposited particles used in the shower reconstruction 
are both larger than ten; 2) the zenith angle of the reconstructed direction is less
than 50$^{\circ}$; 3) the number of particles detected within 40 m from the
shower core is larger than that within $40-100$ m; 4) the shower age is
within $0.6-2.4$. The data are binned into $0.1^{\circ}\times0.1^{\circ}$
pixels in celestial coordinates, and into equal logarithmic energy bins
with bin width $\Delta\log E=0.2$. Due to limited statistics in the outer
Galaxy region, we use a bin width of $\Delta\log E=0.4$.

In this study, the ``direct integral method'' \cite{2004ApJ...603..355F} was 
utilized to estimate the background in each pixel. This approach assumes that 
the collecting efficiency's spatial distribution in the detector coordinates 
remains stable over a short period. The background can be accurately determined 
by convolving the total event rate with the normalized spatial distribution. 
However, due to changes in detector efficiency of EDs caused by temperature 
variations, which differ for different detectors, the spatial distribution undergoes 
slight changes over time during the data taking. We employed a sliding window method 
to obtain a smoothed spatial distribution to address this issue. Specifically, 
a time step of 1 hour was selected, and at each step, events that arrive within 
$\pm5$ hours were used to calculate the spatial distribution. 
This long time window may result in spurious large-scale structures of the
background, which need to be corrected during the analysis (see Sec. A of the
{\tt Supplemental Material}).

To reduce the impacts of known sources to the background estimate, we mask out
relevant sky regions when calculating the background. Detected sources by 
LHAASO-KM2A\footnote{The first LHAASO source catalog can be found in Ref.
\cite{2023arXiv230517030C}. Here, we use each source's fitting location, extension, 
and energy spectrum, assuming a two-dimensional Gaussian morphology and a power-law spectrum.} 
with a pre-trial significance of $5\sigma$, and known sources from those compiled in
TeVCat\footnote{http://tevcat.uchicago.edu} \cite{2008ICRC....3.1341W} were masked 
with mask radii $R_{\rm mask}=n\cdot \sigma$, where $n$ is a constant factor, and
$\sigma=\sqrt{\sigma_{\rm psf}^2+\sigma_{\rm ext}^2}$ is the combined Gaussian
width of the point spread function (PSF) of KM2A $\sigma_{\rm psf}$ and the source 
extension $\sigma_{\rm ext}$. Since the PSF varies with energy, we adopt the largest 
one in the energy bin of $10-15$ TeV to get consistent ROIs for all energy bins. 
For a clean background sky region, $n=5$ is adopted. To avoid repeated masking of 
the KM2A sources and TeVCat ones, whenever the TeVCat source is located within 
$2\sigma$, where $\sigma$ is the Gaussian source width of the KM2A source fitted 
with $E>25$ TeV, they are regarded as the same source and the KM2A parameters are used. 
The Galactic plane with latitude $|b|\leq10^{\circ}$ for declination $\delta\leq 50^{\circ}$ 
and $|b|\leq5^{\circ}$ for $\delta>50^{\circ}$ was also masked. A smaller mask region for 
high declination regions is to ensure sufficient statistics left for an accurate 
background estimate.

\begin{figure*}[!htb]
\centering
\includegraphics[width=1.0\textwidth]{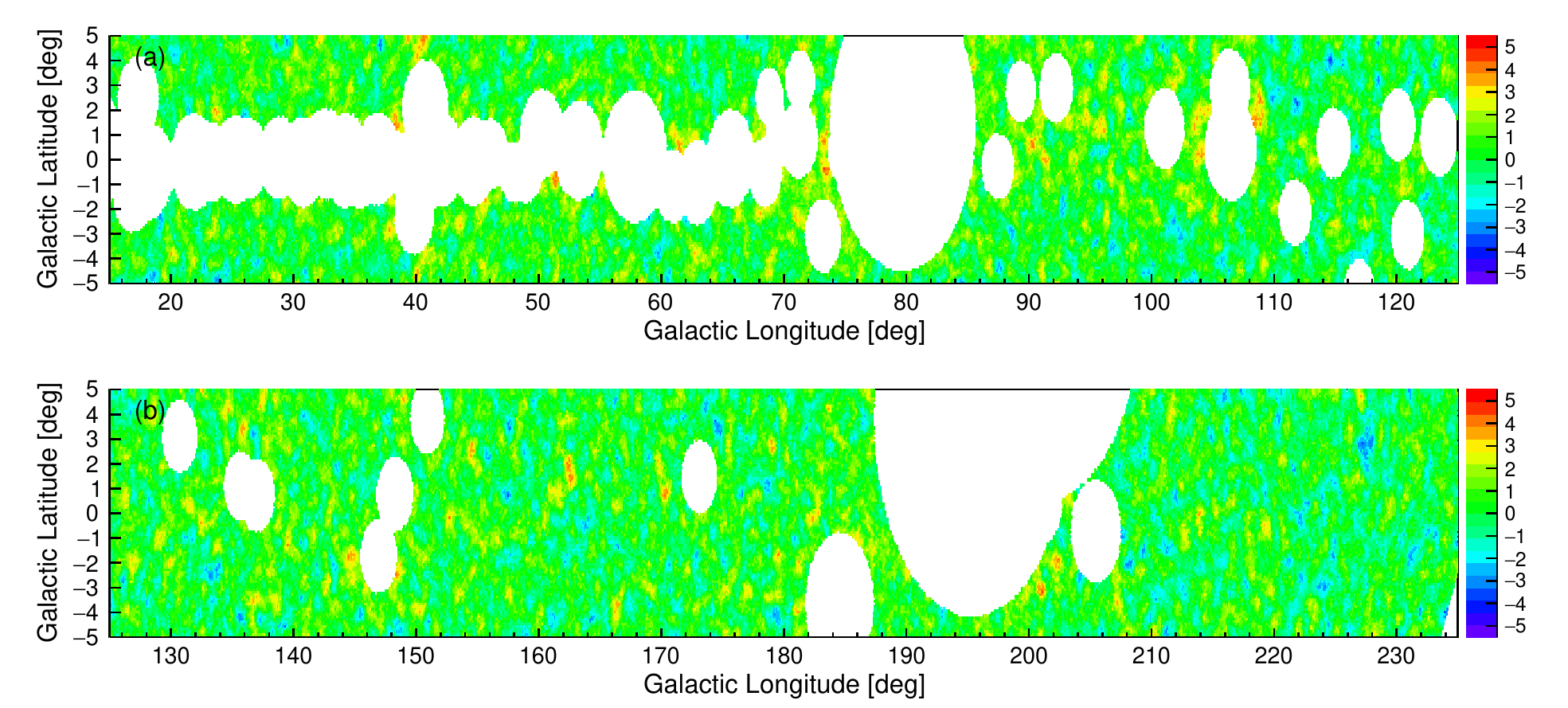}
\caption{The significance maps in Galactic coordinate of the inner Galaxy region 
(panel (a)) and outer Galaxy region (panel (b)) above 25 TeV after masking the 
resolved KM2A and TeVCat sources.}
\label{fig:skymap}
\end{figure*}

To measure the diffuse emission, the contribution from point-like and extended 
sources should be excluded. Similar with above, we mask both the sources detected 
by KM2A and those by other experiments as compiled in TeVCat, but with $n=2.5$ to 
balance the source contamination and the residual sky area. Exceptions are adopted 
for several very extended sources, i.e., $6^{\circ}$ for the Cygnus cocoon and 
$8^{\circ}$ for Geminga and Monogem, which are slightly larger than 2.5 times of
their extensions as compiled in TeVCat. Note that deviations from Gaussian profiles
of these sources may exist \cite{2017Sci...358..911A}.

The residual contamination of resolved sources after the masking is estimated 
from the morphological analysis for both the resolved sources and the diffuse 
emission. We employ the 2D Gaussian templates weighted by the measured fluxes 
for known sources. For the diffuse emission, we use the morphology of the gas 
distribution as traced by the PLANCK dust opacity map, assuming a uniform ratio
between the dust opacity and the gas column \cite{2016A&A...596A.109P}. 
Fitting to the observational data we can obtain the relative contributions of the 
diffuse component and the residual source component. The contamination of resolved 
sources for $n=2.5$ is found to be smaller than 6\% throughout the analyzed 
energy ranges, as summarized in Table S1 of the {\tt Supplemental Material}. 
Due to the improvement of the PSF with energy, the contamination decreases 
efficiently at high energies. The contamination is subtracted when calculating 
the fluxes of the diffuse emission.
%The significance of the diffuse emission is estimated using a test statistic (TS)
%variable defined as two times the logarithmic likelihood ratio, i.e.,
%$TS = 2\ln({\mathcal L}_{s+b}/{\mathcal L}_b)$, where ${\mathcal L}_{s+b}$
%is the likelihood for the signal plus background hypothesis ($H_1$) and ${\mathcal L}_b$ 
%is the likelihood for the background only hypothesis ($H_0$). We assume a power-law model 
%of the spectrum of the diffuse emission, $\phi(E) = \phi_0 \left(E/E_0\right)^{-\alpha}$, 
%where $E_0=50$ TeV is the pivot energy. A forward-folding procedure is then performed 
%to fit the model parameters. Note that we estimate the background from the observational
%data, which has relatively large statistical uncertainty at high energies and needs to be 
%properly accounted for in the likelihood fitting. 

We employ a test statistic (TS) that utilizes twice the logarithmic likelihood 
ratio to determine the significance of the diffuse emission. Specifically, we 
compute $TS = 2\ln({\mathcal L}_{s+b}/{\mathcal L}_b)$, where ${\mathcal L}_{s+b}$ 
and ${\mathcal L}_b$ represent the likelihoods for the signal plus background 
hypothesis ($H_1$) and the background only hypothesis ($H_0$), respectively. 
We assume a power-law model of the spectrum of the diffuse emission in the fitting, 
with $\phi(E)$ expressed as $\phi_0 \left(E/E_0\right)^{-\alpha}$, where $E_0=50$ 
TeV is the pivot energy. We implement a forward-folding procedure to optimize 
the model parameters and estimate the background from the observational data. 
Note that, the statistical uncertainties of the background are relatively large 
at high energies, which need to be properly considered in the fitting process.

The likelihood ratio is defined as
\begin{widetext}
\begin{equation}
\frac{\mathcal{L}_{s+b}}{\mathcal{L}_{b}}=\frac{\prod_{i=0}^{n}{\rm Poisson}\left(N_{i}^{\rm obs},N_{i}^{\rm sig}(\phi_{0},\alpha)+N_{i}^{\rm bkg,1}\right) \cdot {\rm Gauss}\left(N_{i}^{\rm off};N_{i}^{\rm bkg,1},\sigma^{\rm bkg}_i \right)}{\prod_{i=0}^{n}{\rm Poisson}\left(N_{i}^{\rm obs},N_{i}^{\rm bkg,0}\right) \cdot {\rm Gauss}\left(N_{i}^{\rm off};N_{i}^{\rm bkg,0},\sigma^{\rm bkg}_i \right)},
\end{equation}
\end{widetext}
where $N_i^{\rm obs}$ is the observed number of counts in the ROI in the $i$-th 
energy bin, $N_{i}^{\rm off}$ is the estimated background number of counts, 
$N_i^{\rm sig}$ is the predicted number of counts obtained from folding the diffuse 
spectrum to the exposure and response functions (energy and angular) of the KM2A 
detector, $N_{i}^{\rm bkg,0}$ and $N_{i}^{\rm bkg,1}$ are predicted background 
numbers of counts under the hypotheses $H_{0}$ and $H_{1}$, and $\sigma^{\rm bkg}_i$ 
is the statistical uncertainty of the estimated background. Note that 
$N_{i}^{\rm bkg,0}$ and $N_{i}^{\rm bkg,1}$ are nuisance parameters to be fitted.

To determine $\sigma^{\rm bkg}_i$, we generate thousands of mock data sets for 
each energy bin by randomly assigning the arrival time of every event in the 
observational data. We then apply the same background estimation technique to 
each mock data set, which yields a distribution of estimated background counts 
($N_{i}^{\rm off}$) for given energy bin. This distribution can be approximately
described by a Gaussian distribution with width $\sigma^{\rm bkg}_i$.
The likelihood function in Eq. (1) includes a Poisson term, representing the 
statistical probability of the observed number of events, and a Gaussian term, 
representing the probability of the background fluctuation. The flux in each 
energy bin is determined by fitting the normalization parameter $\phi_0$, while 
the spectral index is fixed at the best-fit value obtained from the whole-band fitting.
%The Possion term in the likelihood function is the statistical probability of the
%observed number of events, and the Gauss term is the probability of the estimated 
%background. The flux in each energy bin is obtained through a fitting to the 
%normalization parameter $\phi_0$, with the spectral index being fixed to the best-fit 
%value derived in the whole-band fitting.

{\it Results.} ---
The LHAASO-KM2A significance maps of the two sky regions after masking detected
sources are shown in Fig.~\ref{fig:skymap}. The one-dimensional significance 
distributions are given in Fig.~S2 in the {\tt Supplemental Material}. 
As a comparison, reference regions which are ROIs shifted along the right 
ascension (R.A.) in the celestial coordinates show standard Gaussian 
distributions of the significance, indicating that our background estimate 
is reasonable (Fig.~S2 of the {\tt Supplemental Material}). The total 
significance of the inner (outer) Galaxy region is $29.1\sigma$ ($12.7\sigma$). 
No significant point-like sources are present in the significance maps after 
the mask, except for some hot spots, which need more data to confirm whether 
they are point-like sources or diffuse emissions. The LHAASO results give 
the first measurement of diffuse emission in the outer Galaxy region in the 
VHE-UHE domain.

\begin{figure}[!htb]
\centering
\includegraphics[width=0.48\textwidth]{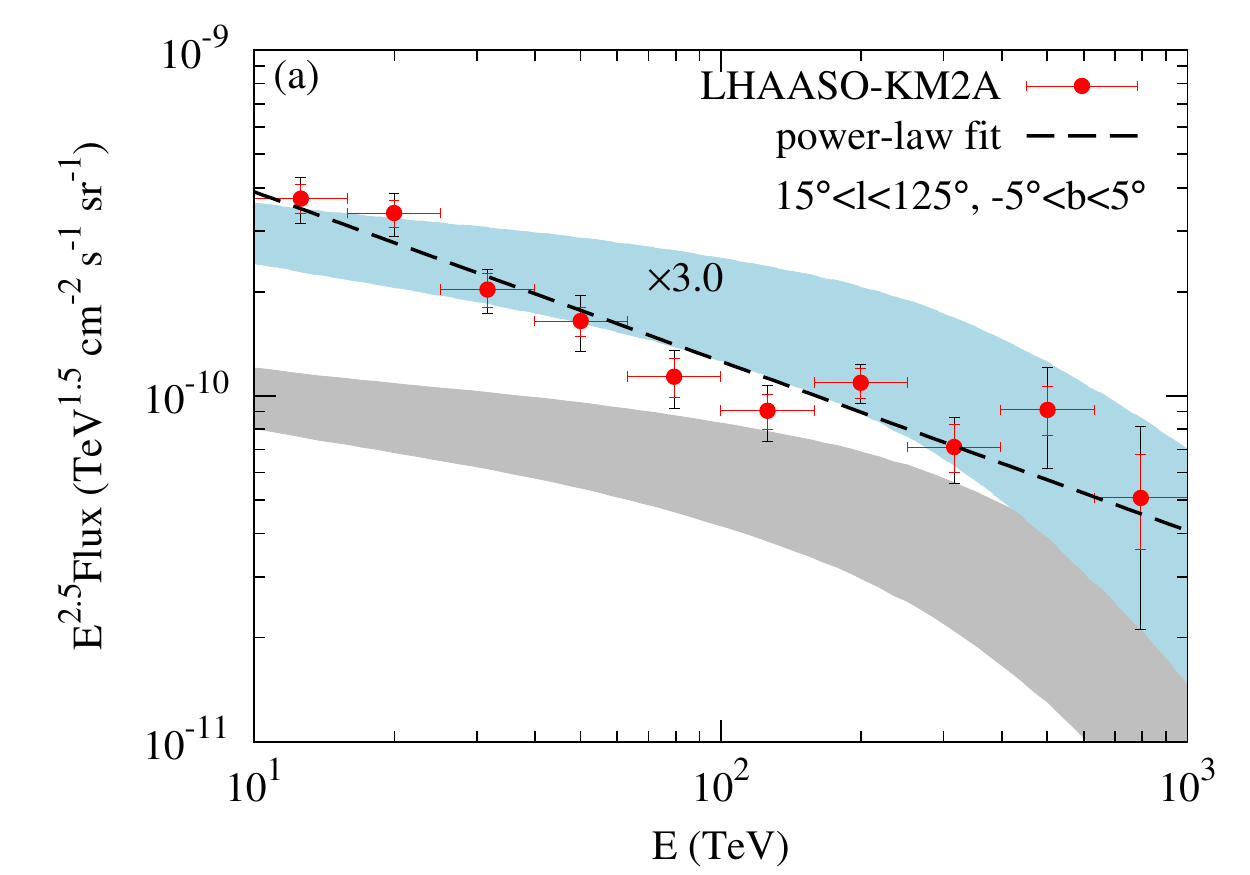}
\includegraphics[width=0.48\textwidth]{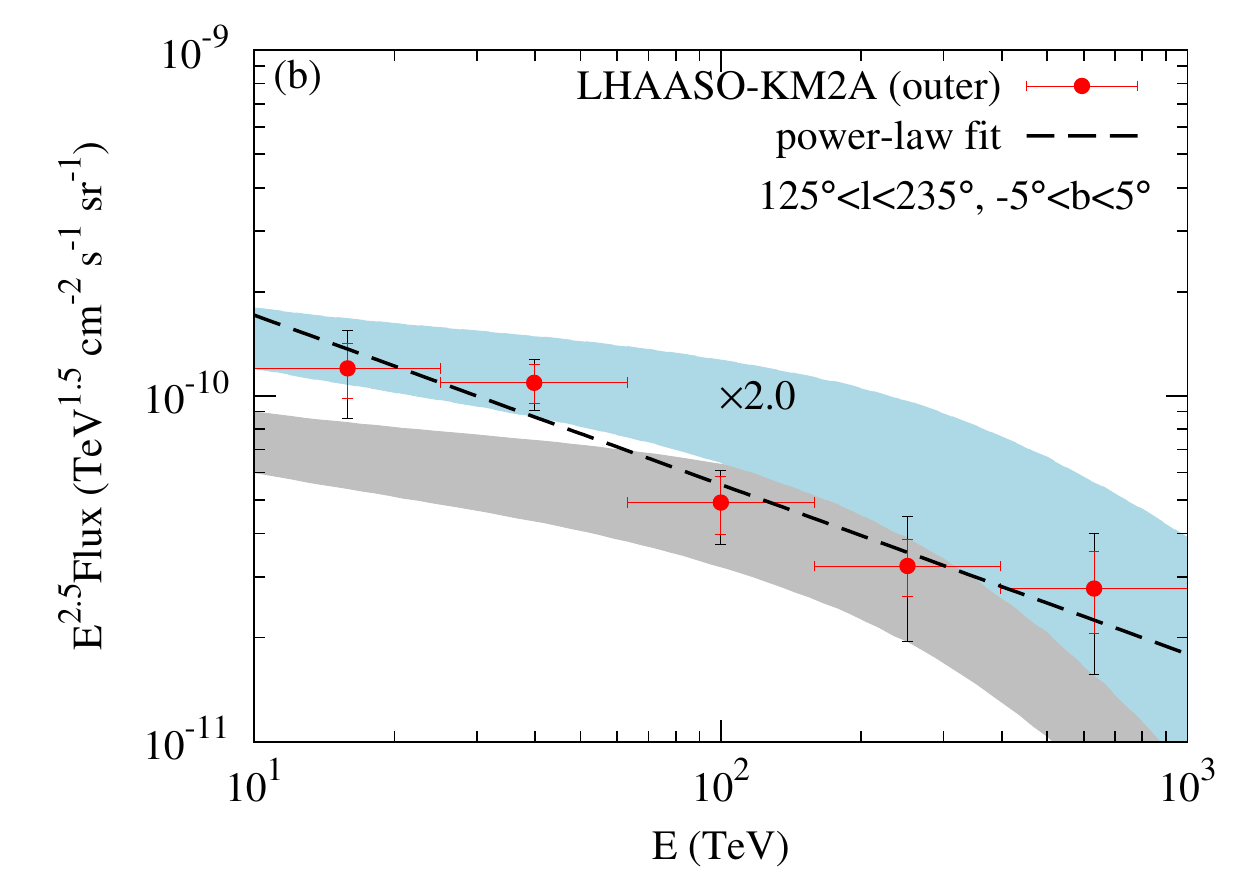}
\caption{Measured fluxes of diffuse $\gamma$-ray emission in the inner and outer Galaxy regions. 
The smaller error bars show the statistical errors and the larger ones show the quadratic
sum of the statistical and systematic errors. In each panel, the dashed line shows the
best-fit power-law function of the data, the grey shaded band shows the model prediction 
assuming local CR spectra and the gas column density with the same mask as the data, 
and the cyan shaded band is the grey one multiplied by a constant factor of 3.0 for the
inner region and 2.0 for the outer region.}
\label{fig:spec}
\end{figure}

Fig.~\ref{fig:spec} shows the derived fluxes of the diffuse emission in the
two regions. The fluxes in different energy bins are tabulated in Tables 
S2 and S3 of the {\tt Supplemental Material}).
%For comparison, the previous measurement by Tibet-AS$\gamma$ in the same sky 
%region is also shown in Fig.~\ref{fig:spec} \cite{2021PhRvL.126n1101A}.
%Note that the source mask radii and the source list of the Tibet-AS$\gamma$ 
%analysis are different from the present one, and thus quantitative comparison 
%among these measurements should be done with caution. 
From Fig.~\ref{fig:skymap} we can see that considerable regions along the 
innermost Galactic disk are masked for the inner Galaxy region. Since the 
expected diffuse emission is non-uniform, the current measurements are thus 
not equivalent to the total average emission in the ROIs. As an estimate, we 
find that the average diffuse emission in the ROIs without any masking will 
be higher by $\sim61\%$ and $\sim2\%$ than our measurements assuming a spatial
template of the PLANCK dust opacity map in the inner and outer Galactic regions,
respectively.

We fit the measured spectrum using a power-law function, finding that the index is 
$-2.99\pm0.04_{\rm stat}$ for the inner Galaxy region and $-2.99\pm0.07_{\rm stat}$ 
for the outer Galaxy region (see Table \ref{table:fitspec}). Possible spectral 
structures deviating from power-laws are not significant, and more data statistics 
are needed to further address such issues.
As a comparison, the power-law fitting to the spectrum without subtracting
the residual source contamination as given in Table S1 obtains $-3.01\pm0.04_{\rm stat}$ 
for the inner region and $-2.99\pm0.07_{\rm stat}$ for the outer region, indicating
that the effect due to residuals of known sources is minor.

\begin{table}[!htb]
\centering
\caption {Fitting parameters of the LHAASO-KM2A diffuse spectra.}
\begin{tabular}{lccc}
\hline \hline
 & $\phi_0$ & $\alpha$ \\
 & ($10^{-14}~{\rm TeV^{-1}~cm^{-2}~s^{-1}~sr^{-1}}$) &  \\
\hline
Inner Galaxy & $1.00\pm0.04_{\rm stat}\pm0.09_{\rm sys}$ & $-2.99\pm0.04_{\rm stat}\pm0.07_{\rm sys}$ \\
Outer Galaxy & $0.44\pm0.04_{\rm stat}\pm0.05_{\rm sys}$ & $-2.99\pm0.07_{\rm stat}\pm0.12_{\rm sys}$ \\
\hline
\hline
\end{tabular}
\label{table:fitspec}
\end{table}

\begin{figure*}[!htb]
\centering
\includegraphics[width=1.0\textwidth]{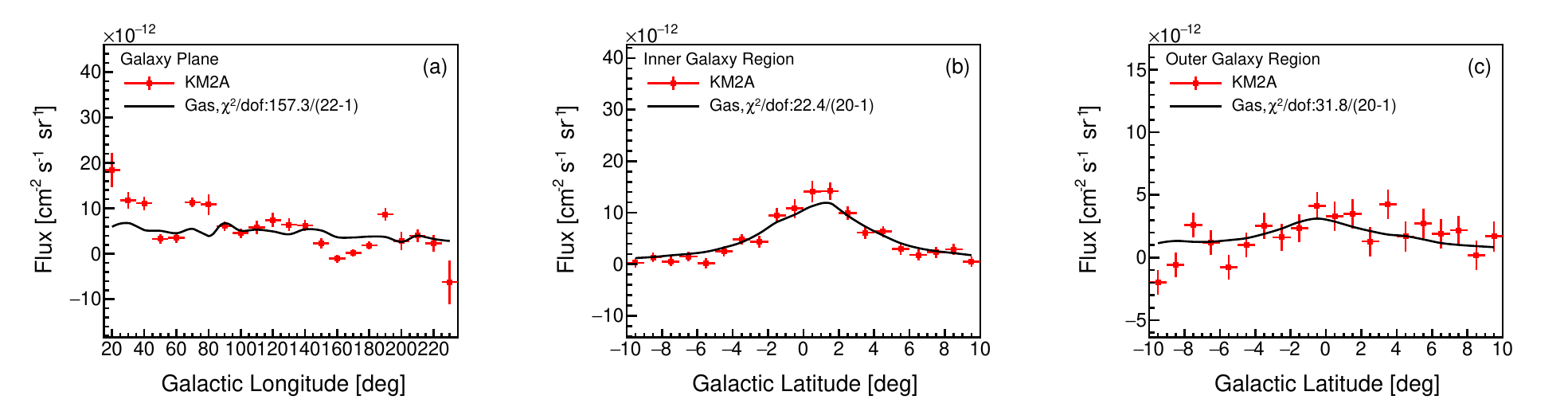}
\includegraphics[width=1.0\textwidth]{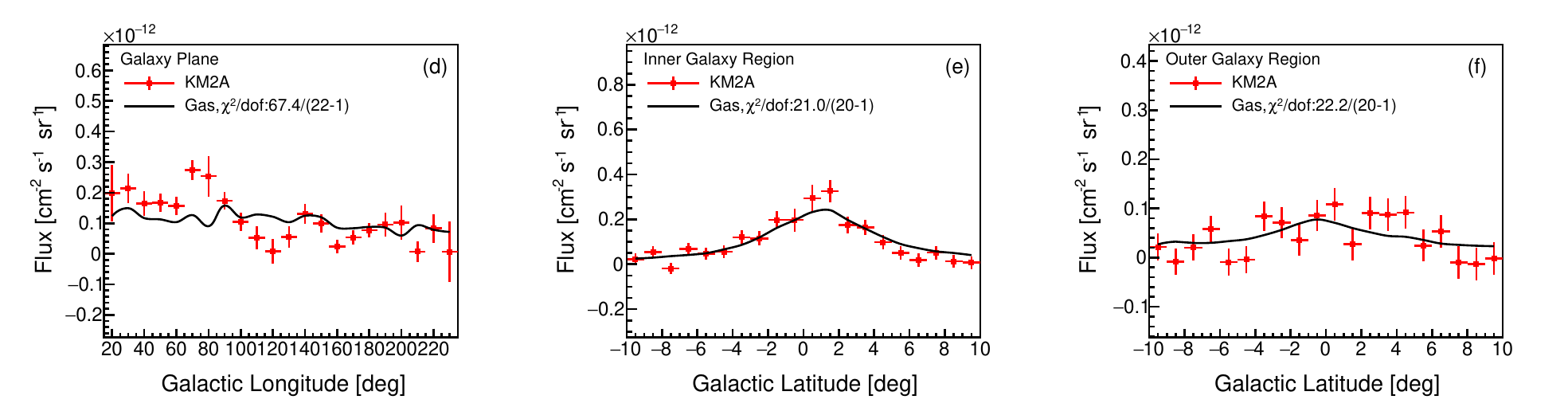}
\caption{Galactic longitude and latitude profiles of the diffuse emission for 
energy bands $10-63$ TeV (top three panels) and $63-1000$ TeV (bottom three 
panels), respectively. The solid line in each panel is the best-fit gas
distribution according to the PLANCK dust opacity map.}
\label{fig:1d_prof}
\end{figure*}

In Fig.~\ref{fig:1d_prof}, we present the longitude and latitude
profiles for the two sky regions, for energy bands of $10-63$ TeV and
$63-1000$ TeV. The latitude integration range when deriving the longitude
profile is from $-5^{\circ}$ to $+5^{\circ}$, and the longitude integration
ranges for the latitude profiles are the same as the definitions of the ROIs.
The diffuse emission shows a clear decrease from the inner Galaxy to the
outer Galaxy and a concentration in the low Galactic latitudes.
We fit the longitude and latitude distributions using the gas template traced 
by the PLANCK dust opacity map, as shown by the solid line in each panel. 
The results show that the measured latitude distributions generally agree 
with the gas distribution, except for a slight deviation for $10-63$ TeV
profile in the outer region (the $p$-value of the fitting is about 0.03). 
We can see a clear deviation of the data from the gas template for the 
longitude distribution. The fitting gives $\chi^2/dof=157.3/21$ and 67.4/21 
for $10-63$ TeV and $63-1000$ TeV energy bands, corresponding to $p$-values 
of about $7\times10^{-23}$ and $10^{-6}$, respectively. 
The results indicate that the gas distribution may not well trace the 
diffuse $\gamma$-ray emission at very high energies. We calculate the 
angular power spectrum of the relative $\gamma$-ray flux map with $E>25$ 
TeV, and find that it is consistent with the angular power spectrum of 
the gas distribution for multipole $l>10$ but shows slightly higher power 
for smaller $l$, which may indicate that the data are more clumpy than 
the gas distribution. See Fig.~S3 of the {\tt Supplemental Material}.
We also fit the latitude profiles by adding a Gaussian latitude distribution 
centered at $b=0$ to the gas template but find only slight improvements in 
the goodness-of-fit (see Fig.~S4 of the {\tt Supplemental Material}).

{\it Systematic uncertainties.} ---
The event rate varies during the operation due to the variation of temperature 
and humidity of the atmosphere, affecting the detection efficiency for 
$\gamma$-rays. This effect results in about 7\% systematic uncertainty for 
the flux ($\phi_0$) and 0.02 for the spectral index ($\alpha$) for point-like 
sources \cite{2021ChPhC..45b5002A}. For the diffuse emission in this work, 
we expect that the variation of atmospheric conditions contributes to similar 
systematic uncertainties since its main impact is on the detection efficiency. 
The array layout changed slightly for debugging purposes during the operation, 
which results in about $1\%$ variation for $\phi_0$ and 0.02 change for 
$\alpha$, estimated from simulations with two layouts. To account for the 
systematic uncertainties from the background estimate method, we vary the 
time window for background estimate from $\pm5$ hours to $\pm2$, $\pm6$, 
and $\pm12$ hours, 
%vary the time coordinate from the Local Sidereal Time to the Modified Julian Date, 
vary the mask maps for the background estimate (e.g., $|b|\leq 5^{\circ}$ for 
declination $\delta\leq 60^{\circ}$ which enables shorter time windows), and 
test different large-scale efficiency correction parameters, and obtain the 
impacts on $\phi_0$ of about $5\%~(10\%)$ and on $\alpha$ of about 0.05 (0.10), 
for the inner (outer) region.

The $\gamma$-ray survival fraction as a function of the $\gamma$/CR 
discrimination parameter $R_{\mu e}$ is obtained by Monte Carlo (MC) 
simulations. The difference in the survival fraction between MC and 
experimental data may lead to systematic uncertainties. We compared the 
spectra using different groups of $R_{\mu e}$, and estimated the systematic 
uncertainties to be about 2\% (5\%) for $\phi_0$ and 0.04 (0.06) for $\alpha$ 
for the inner (outer) Galaxy region. By combining all these systematic 
uncertainties in quadrature, the overall systematic uncertainties are 
9\% (12\%) for $\phi_0$ and 0.07 (0.12) for $\alpha$, for the inner 
(outer) region. The systematic uncertainties of the flux in each energy 
bin are given in Tables S2 and S3 of the {\tt Supplemental Material}.

{\it Discussion.} ---
We compare the LHAASO measurements of diffuse emission with the predictions
of hadronic interactions between CRs and the ISM. While the CR spectra were 
directly measured with relatively small uncertainties below $\sim100$ TeV 
\cite{2009BRASP..73..564P,2017ApJ...839....5Y,2019SciA....5.3793A,2021PhRvL.126t1102A}, 
the uncertainties at higher energies are large, particularly for individual 
elements \cite{2013APh....47...54A,2019PhRvD.100h2002A}. We use the sum of two 
power-law functions with an exponential cutoff, $\sum_{i=1,2} A_iE^{-B_i}\exp(-E/C_i)$, 
to describe the local spectra of both protons and helium nuclei, and adjust 
the parameters $A_i,\ B_i,\ C_i$ to give low and high fittings to the data 
(see Fig.~S5 and Table S4 of the {\tt Supplemental Material}). 
Assuming the CR intensity is uniform in the Galaxy as a zero-order approximation, 
we obtain the expected diffuse $\gamma$-ray emission in the two regions from hadronic 
interactions between CR nuclei and the ISM. We assume the ISM hygrogen and 
helium distributions follow the PLANCK dust opacity map \cite{2016A&A...596A.109P}, 
and the number density of helium to hydrogen is $1:10$. 
Heavier nuclei are expected to contribute to a fraction of $\lesssim10\%$ of the 
total emission and are neglected in this work \cite{2022A&A...661A..72B}.
The secondary $\gamma$ production spectrum is calculated using the {\tt AAfrag} 
package \cite{2019CoPhC.24506846K}. We also include the $\gamma\gamma$ absorption
effect\footnote{The absorption optical depth is a function of photon energy
and travel path, and thus depends on the spatial emissivity of the $\gamma$-ray 
radiation. We employ the GALPROP code \cite{2000ApJ...537..763S,2004ApJ...613..962S}
to calculate the diffuse $\gamma$-ray emissivity and implement the absorption
obtained with the ISRF embedded in GALPROP, using propagation parameters 
determined by recent CR observations \cite{2020JCAP...11..027Y}.},
which is important for $E>100$ TeV \cite{2006A&A...449..641Z}.

The results are given by grey shaded bands in Fig.~\ref{fig:spec}, 
with bandwidth representing the uncertainty from the CR flux measurements.
It can be seen that the predicted fluxes are lower than the measured
ones\footnote{Note that the spatial inhomogeneity of CR intensity may change 
the prediction. As an estimate, for a GALPROP model with parameters given in
Ref.~\cite{2020JCAP...11..027Y}, including the spatial distribution of CRs 
results in 5\% (41\%) lower fluxes in the inner (outer) region.}.
As a comparison, we scale the grey bands by a factor of 3 for the inner region 
and 2 for the outer region, as shown by the cyan shaded bands, which can 
roughly match with the data.
The spectral shapes of the measurements are basically power-laws, which 
are slightly different from the predicted curved spectra.
To quantify their differences, we fit the fluxes (with statistical and 
systematic uncertainties added in quadrature) with the predicted ``high" 
spectral shape (with normalization free) which gives $\chi^2/dof=33.8/9$ 
(7.7/4) in the inner (outer) region, and with the ``low" spectral shape 
which gives $\chi^2/dof=21.9/9$ (4.4/4). The $p$-values of the fittings 
are about $10^{-4}\sim 10^{-2}$ for the inner region, and $0.10\sim0.35$ 
for the outer region. As a comparison, the power-law fitting gives 
$\chi^2/dof=9.1/8$ (2.2/3).
For the inner region, excesses exist for the whole energy band, while for the 
outer region, excesses mainly appear at low energies (for $E\lesssim 60$ TeV).
The LHAASO measurements may thus imply the existence of additional sources 
for the diffuse emission. The excesses may come from unresolved sources.
Although sources above $5\sigma$ detected by KM2A and other experiments were 
masked from the data, there should be sources below the chosen threshold 
whose accumulative contribution may account for a fraction of the observed 
emission \cite{2020A&A...643A.137S}. The ICS from electrons and positrons 
injected from pulsars or pulsar wind nebulae may be proper candidates \cite{2000A&A...362..937A,2004vhec.book.....A,2018PhRvL.120l1101L,
2022ApJ...928...19V}. Some such sources were detected as extended 
halos in the VHE band \cite{2017Sci...358..911A,2021PhRvL.126x1103A}.
It is expected that such pulsar halos may be general in the Milky Way,
giving remarkable diffuse emission in the VHE to UHE band
\cite{2018PhRvL.120l1101L,2022ApJ...928...19V}. The imperfect correlation
of the longitudinal distributions of the data with gas may support such
a scenario. Alternatively, CR interactions with the medium around acceleration 
sources \cite{2019PhRvD.100f3020Y,2022PhRvD.105b3002Z}, the spatial variations 
of the CR spectra (e.g., harder spectra in places other than the local vicinity;
\cite{2018PhRvD..97f3008G,2018PhRvD..98d3003L,2023A&A...672A..58D}) or 
dust-to-gas ratio \cite{2017A&A...606L..12G} may also explain the excesses.

{\it Conclusion.}---
We report the measurements of the diffuse $\gamma$-ray emission in the VHE to UHE
window of $10-1000$ TeV from the Galactic plane using the LHAASO-KM2A data.
In total, 302 days of the half array data, 219 days of the 3/4 array data, 
and 423 days of the full array data of LHAASO-KM2A are used in this work. 
To reduce the contamination from detected sources, the sky regions around 
known VHE/UHE sources and those newly detected by the LHAASO-KM2A are masked. 
In the Galactic plane, Two sky regions (Fig.~\ref{fig:skymap}), the inner 
Galaxy and the outer Galaxy regions, are analyzed. After masking the sources, 
we find significant diffuse emission above 10 TeV  with $29.1\sigma$ and 
$12.7\sigma$ significance for the inner and outer Galaxy regions, respectively. 
The outer Galaxy region is, for the first time, to be observed to have VHE-UHE 
diffuse emission. A power-law can well describe the spectra in both the inner 
and outer regions with similar spectral indices of $-2.99$. Compared with the 
prediction of CR interactions with the ISM, the LHAASO measured fluxes are 
higher by a factor of $2\sim3$ in both regions (for the outer region the 
excess is evident for $E\lesssim60$ TeV). The latitude distributions of the 
diffuse emission are in general consistent with the gas distribution, 
while deviation is shown in the longitude distribution. The KM2A measurements 
provide interesting insights in probing the source distribution and interactions 
of CRs in the Galaxy. Further understanding of the nature of the diffuse emission 
is expected to be achieved with the accumulation of more data by KM2A and the 
analysis from sub-TeV to $\sim10$ TeV energies with the WCDA data, and/or a joint 
modeling of the recently reported detection of neutrino emission from the Galactic 
plane by IceCube \cite{IceCube2023}.

\acknowledgements
We would like to thank all staff members who work at the LHAASO site above 4400 
meters above sea level year-round to maintain the detector and keep the water 
recycling system, electricity power supply and other components of the experiment 
operating smoothly. We are grateful to Chengdu Management Committee of Tianfu New 
Area for the constant financial support for research with LHAASO data. 
This work is also supported by the following grants: the National Key R\&D program 
of China under grants 2018YFA0404202, 2018YFA0404201, 2018YFA0404203, and 2018YFA0404204, 
the National Natural Science Foundation of China under grants 12220101003, 12273114, 
12022502, 12205314, 12105301, 12261160362, 12105294, U1931201, the Project for Young 
Scientists in Basic Research of Chinese Academy of Sciences under grant YSBR-061, 
the Program for Innovative Talents and Entrepreneur in Jiangsu,
and the NSRF of Thailand via the Program Management Unit for Human Resources \& 
Institutional Development, Research and Innovation under grant B37G660015.

\bibliographystyle{apsrev}
\bibliography{refs}

\clearpage
\setcounter{figure}{0}
\renewcommand\thefigure{S\arabic{figure}}
\setcounter{table}{0}
\renewcommand\thetable{S\arabic{table}}

\section*{Supplemental Material}

\subsection{Correction of large-scale structures}

\begin{figure}[!htb]
\centering
\includegraphics[width=0.48\textwidth]{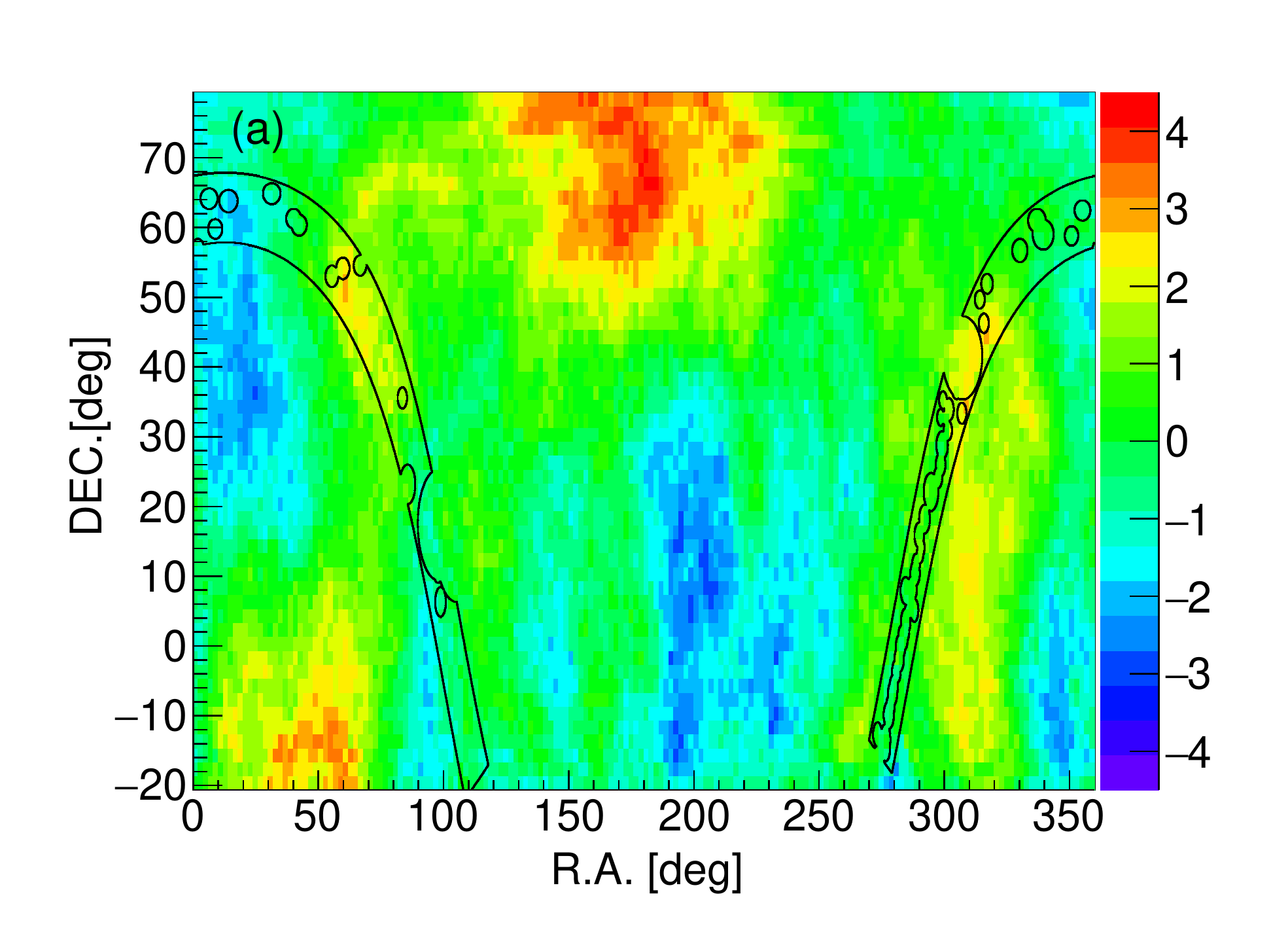}
\includegraphics[width=0.48\textwidth]{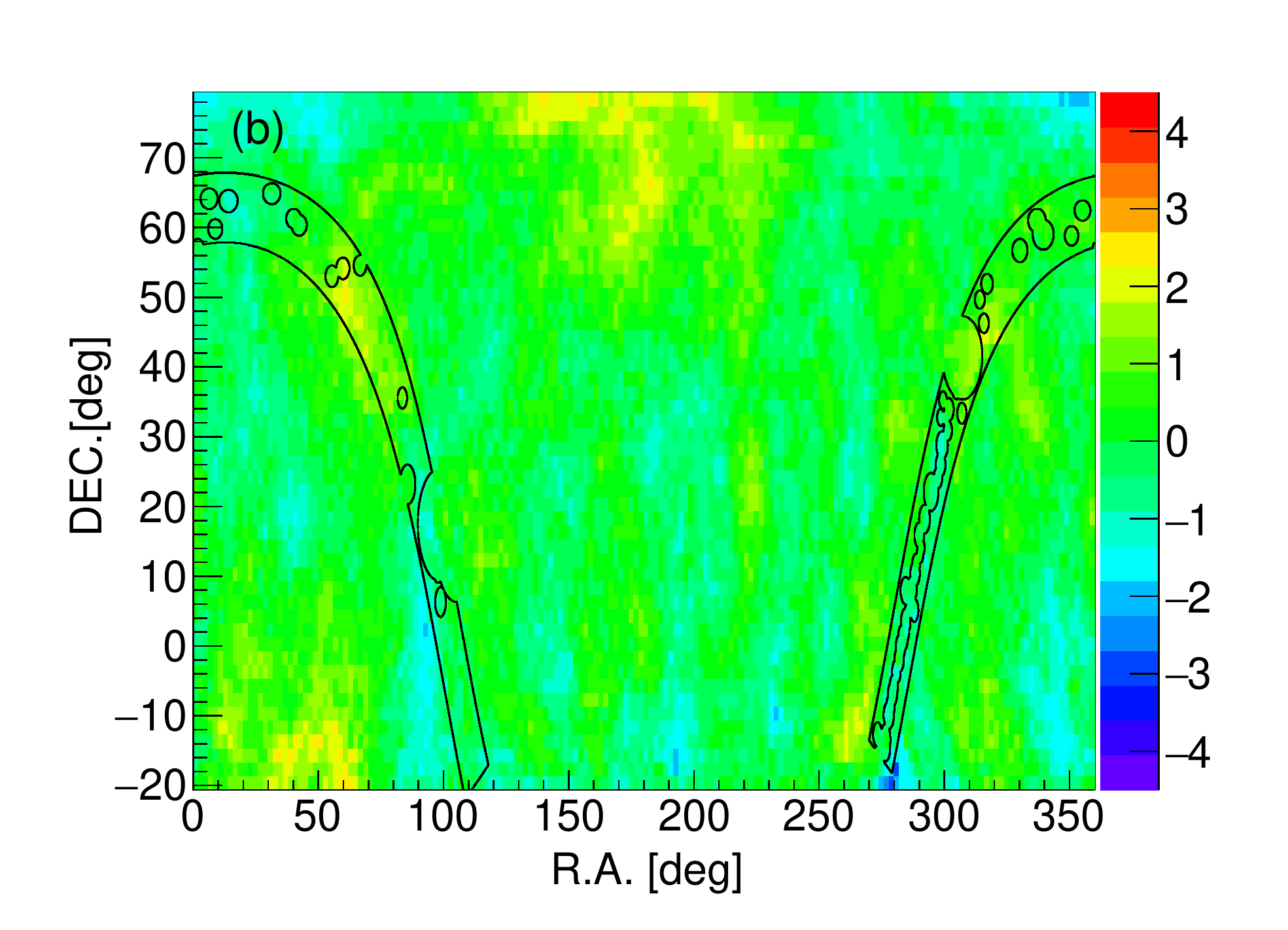}
\caption{Significance maps in celestial coordinate for energy bin of $10-15$ TeV 
for the KM2A full array data. Panel (a) and (b) represent the results before and 
after the large-scale structure correction, respectively. The black solid lines 
show the ROIs of this analysis.}
\label{fig:LargeScale}
\end{figure}

The ``direct integration method'' of the background estimate may introduce 
large-scale structures using events within a relatively long time window 
(10 hours), which need to be corrected. The possible origin of such structures 
can be attributed to two causes: the true large-scale anisotropy of $\gamma$-like 
CR events and the spurious anisotropy due to variations in the spatial 
distribution of the efficiency over the time window. To correct the large-scale 
structures, we smooth the ON and OFF counts map (with known sources and the 
Galactic plane, $|b|\leq10^{\circ}$ for declination $\delta\leq 50^{\circ}$ 
and $|b|\leq5^{\circ}$ for $\delta>50^{\circ}$, masked) with $20^{\circ}$ 
radius top-hat kernel, and then correct the estimated 
background by multiplying $N_{\rm on}/N_{\rm off}$ of each pixel obtained 
after the smoothing. The resulting significance map shows no such large-scale
structures any more, as shown in Fig.~\ref{fig:LargeScale}. 
Different smoothing window sizes are tested, and the changes of the results are
taken as systematic uncertainties.

\subsection{Residual fractions from resolved sources}

Table \ref{table:residual} gives the calculated residual fractions of the signal
that is from resolved sources, due to the tails of the extension of sources. 
In this work, the source residuals have been subtracted from the signal to obtain 
the diffuse emission fluxes. 

\begin{table}[!htb]
\centering
\caption{Percentage contribution of resolved sources to the total signal.}
\begin{tabular}{ccc}
\hline
$\log(E_{\rm rec}/$TeV) & Inner Galaxy (\%) & Outer Galaxy (\%) \\ 
\hline
1.0-1.2 & $5.97 \pm 0.67$ & $4.58 \pm 1.63$ \\ 
1.2-1.4 & $4.26 \pm 0.43$ & $2.25 \pm 0.44$ \\ 
1.4-1.6 & $2.97 \pm 0.36$ & $1.39 \pm 0.23$ \\ 
1.6-1.8 & $1.95 \pm 0.21$ & $1.88 \pm 0.45$ \\ 
1.8-2.0 & $1.97 \pm 0.27$ & $0.77 \pm 0.16$ \\ 
$>2.0$  & $0.76 \pm 0.06$ & $0.39 \pm 0.09$ \\ 
\hline
\end{tabular}
\label{table:residual}
\end{table}

\subsection{One-dimensional distributions of ROIs and control regions}
\begin{figure*}[!htb]
\centering
\includegraphics[width=0.8\textwidth]{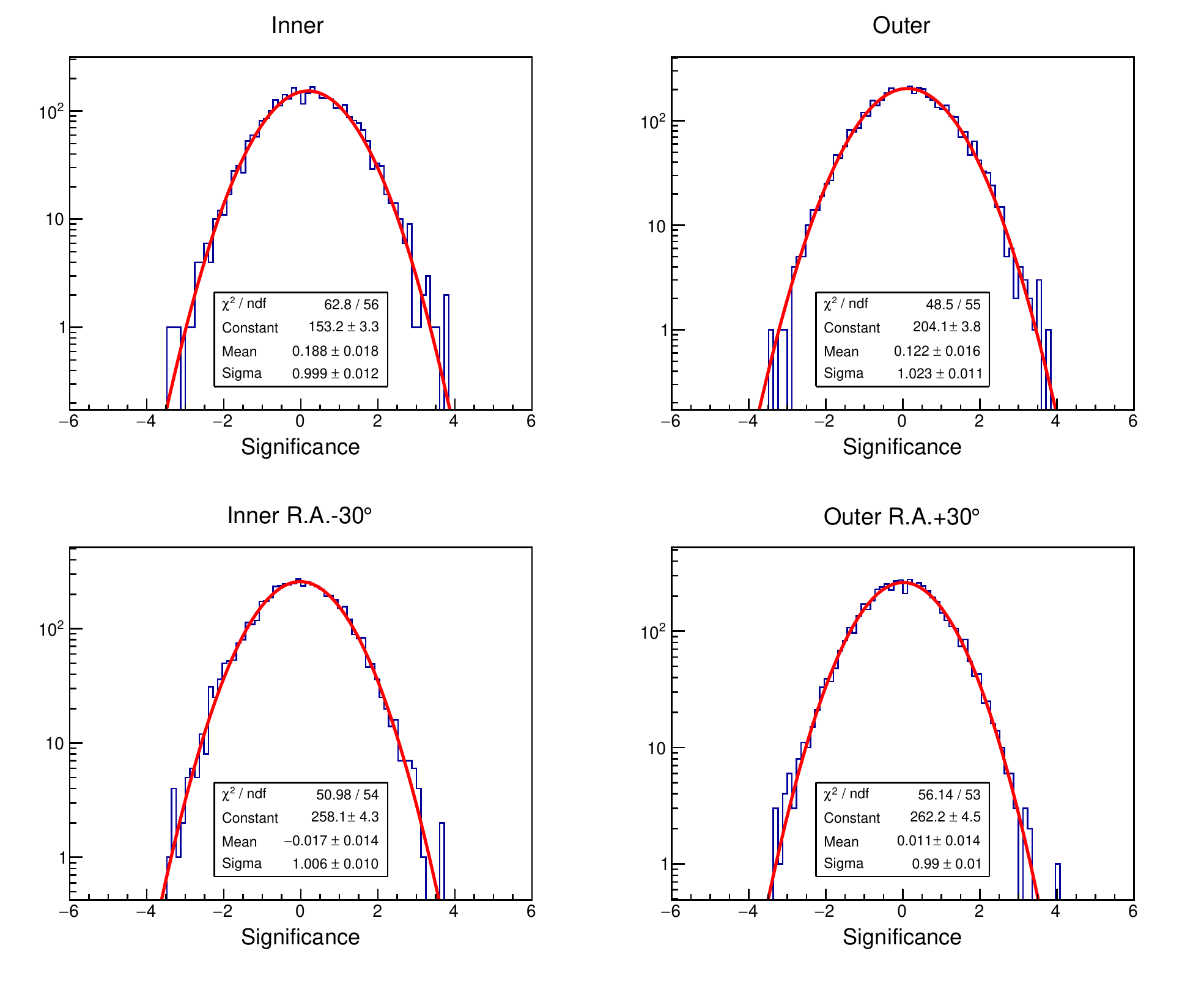}
\caption{One-dimensional significance distributions above 25 TeV after source masks. 
The pixel size is $0.5^{\circ}\times0.5^{\circ}$. Top two panels are for our ROIs, 
and bottom panels are the reference sky regions shifted to lower R.A. by 30 degrees 
for the inner Galaxy region and higher R.A. by 30 degrees for the outer Galaxy region.}
\label{fig:1d_sig}
\end{figure*}

Fig.~\ref{fig:1d_sig} shows the one-dimensional significance distributions of 
our analyzed ROIs (top two panels) and the control sky regions with ROIs shifted
along the R.A. for $-30^{\circ}$ and $+30^{\circ}$. For our ROIs, we see that the 
mean values are positive, suggesting the existence of diffuse emission in these
regions. The significance distributions of the control regions are consistent with 
the standard Gaussian distribution expected for the background distribution.

\subsection{Fluxes of diffuse emission measured by LHAASO-KM2A}

Tables \ref{table:flux_in} and \ref{table:flux_out} present the fluxes of the 
diffuse emission in the inner and outer Galaxy regions measured by KM2A, with
$1\sigma$ statistical errors and the estimated systematic uncertainties.

\begin{table}[!htb]
\centering
\caption{Fluxes with $1\sigma$ statistical uncertainties and systematic uncertainties 
of the diffuse emission in the inner Galaxy region measured by KM2A.}
\begin{tabular}{cccc}
\hline
$\log(E_{\rm rec}/$TeV) & $\langle E\rangle$ & $\phi\pm\sigma_{\rm stat}\pm\sigma_{\rm sys}$ \\
 & (TeV) & (TeV$^{-1}$~cm$^{-2}$~s$^{-1}$~sr$^{-1}$) \\
\hline
%1.0-1.2 & 12.6  & $(6.85 \pm 0.65 \pm 0.94) \times10^{-13}$\\ 
%1.2-1.4 & 20.0  & $(1.94 \pm 0.18 \pm 0.18) \times10^{-13}$\\ 
%1.4-1.6 & 31.6  & $(3.48 \pm 0.41 \pm 0.43) \times10^{-14}$\\ 
%1.6-1.8 & 50.1  & $(9.69 \pm 0.93 \pm 1.96) \times10^{-15}$\\ 
%1.8-2.0 & 79.4  & $(2.17 \pm 0.28 \pm 0.36) \times10^{-15}$\\ 
%2.0-2.2 & 125.9 & $(6.05^{+0.66}_{-0.65} \pm 0.74) \times10^{-16}$\\ 
%2.2-2.4 & 199.5 & $(2.16^{+0.21}_{-0.21} \pm 0.19) \times10^{-16}$\\ 
%2.4-2.6 & 316.2 & $(4.38^{+0.70}_{-0.67} \pm 0.57) \times10^{-17}$\\ 
%2.6-2.8 & 501.2 & $(1.64^{+0.28}_{-0.26} \pm 0.35) \times10^{-17}$\\
%2.8-3.0 & 794.3 & $(2.68^{+0.95}_{-0.83} \pm 1.17) \times10^{-18}$\\
1.0-1.2& 12.6  &  $(6.61 \pm 0.64 \pm 0.77) \times 10^{-13}$\\ 
1.2-1.4&20.0    & $(1.90 \pm 0.17 \pm 0.21) \times 10^{-13}$ \\ 
1.4-1.6&31.6  &  $(3.61 \pm 0.40 \pm 0.34) \times 10^{-14}$ \\ 
1.6-1.8&50.1  &  $(9.26 \pm 0.90 \pm 1.47) \times 10^{-15}$  \\ 
1.8-2.0&79.4  &  $(2.02 \pm 0.26 \pm 0.29) \times 10^{-15}$  \\ 
2.0-2.2&125.9 & $(5.09 ^{+0.60}_{-0.59} \pm 0.72) \times 10^{-16}$\\ 
2.2-2.4&199.5 & $(1.94 ^{+0.19}_{-0.19} \pm 0.16) \times 10^{-16}$\\ 
2.4-2.6&316.2 &  $(4.00 ^{+0.64}_{-0.62} \pm 0.60) \times 10^{-17}$ \\ 
2.6-2.8&501.2 &  $(1.62 ^{+0.27}_{-0.25} \pm 0.46) \times 10^{-17}$ \\
2.8-3.0&794.3&  $(2.85 ^{+0.95}_{-0.82} \pm 1.45) \times10^{-18}$ \\
\hline
\end{tabular}
\label{table:flux_in}
\end{table}

\begin{table}[!htb]
\centering
\caption{Fluxes with $1\sigma$ statistical uncertainties and systematic uncertainties 
of the diffuse emission in the outer Galaxy region measured by KM2A.}
\begin{tabular}{cccc}
\hline
$\log(E_{\rm rec}/$TeV) & $\langle E\rangle$ & $\phi\pm\sigma_{\rm stat}\pm\sigma_{\rm sys}$ \\
 & (TeV) & (TeV$^{-1}$~cm$^{-2}$~s$^{-1}$~sr$^{-1}$) \\
\hline
%1.0-1.4 & 15.8  & $(1.30 \pm 0.23 \pm 0.33) \times10^{-13}$ \\ 
%1.4-1.8 & 39.8  & $(1.24 \pm 0.14 \pm 0.23) \times10^{-14}$ \\ 
%1.8-2.2 & 100.0 & $(5.94 \pm 1.02 \pm 0.92) \times10^{-16}$ \\ 
%2.2-2.6 & 251.2 & $(3.12^{+0.66}_{-0.65} \pm 1.49) \times10^{-17}$\\ 
%2.6-3.0 & 631.0 & $(2.40^{+0.79}_{-0.73} \pm 0.65) \times10^{-18}$ \\ 
1.0-1.4& 15.8 &	$(1.20 \pm 0.22 \pm 0.26) \times10^{-13}$ \\ 
1.4-1.8 & 39.8& $(1.09 \pm 0.14 \pm 0.12) \times10^{-14}$ \\ 
1.8-2.2 & 100.0&	$(4.91\pm 0.94 \pm 0.74) \times10^{-16}$ \\ 
2.2-2.6 &251.2&	$(3.22 ^{+0.61}_{-0.59} \pm 1.12) \times10^{-17}$ \\ 
2.6-3.0&631.0&	$(2.77 ^{+0.77}_{-0.72} \pm 0.97) \times10^{-18}$ \\ 
\hline
\end{tabular}\\
\label{table:flux_out}
\end{table}

\subsection{Angular power spectrum}

\begin{figure}[!htb]
\centering
\includegraphics[width=0.48\textwidth]{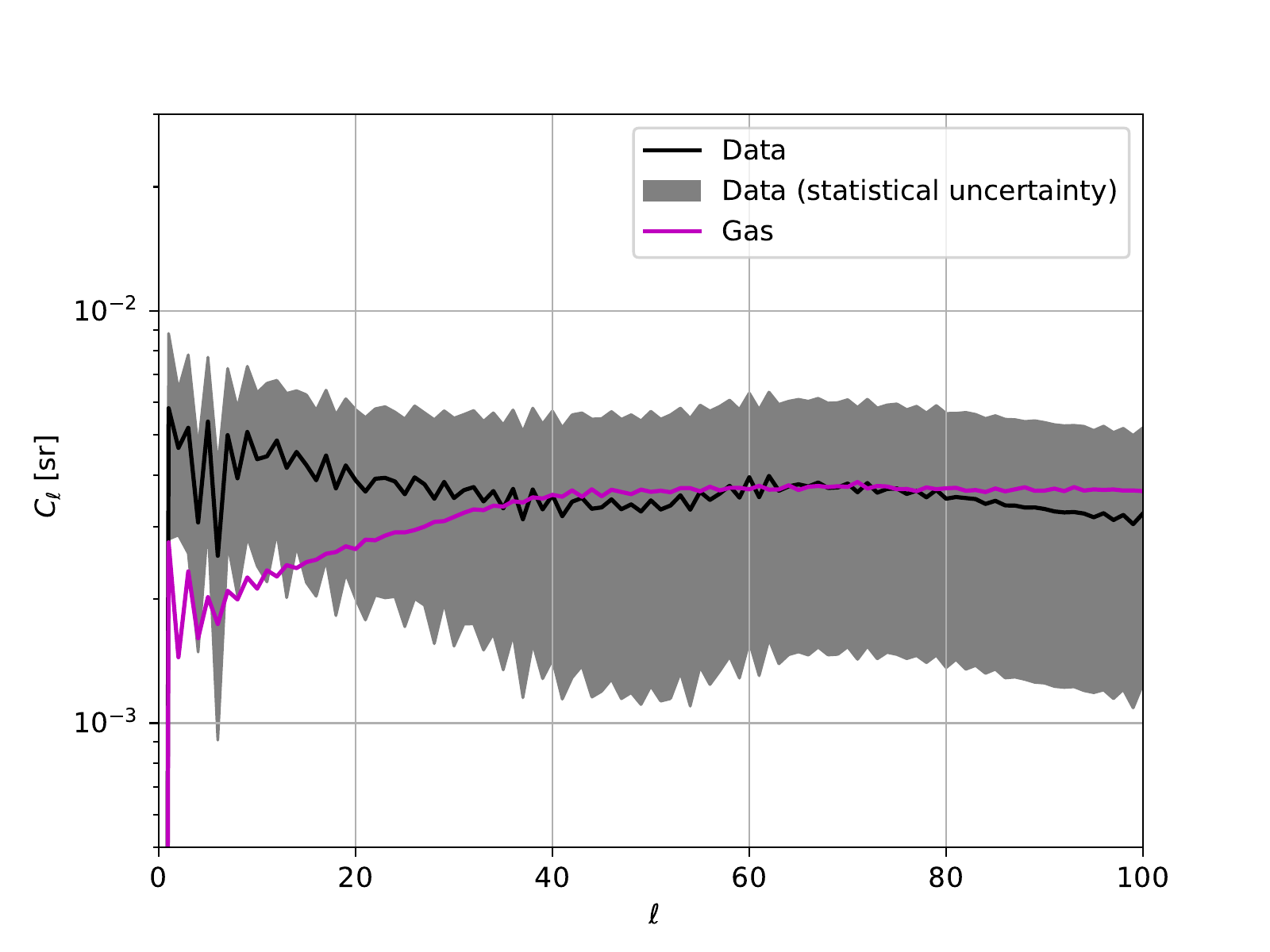}
\caption{Angular power spectra of the data (black) and the $1\sigma$ uncertainty band 
(gray shaded band), compared with the expectation based on the PLANCK-derived gas 
distribution (purple). The same mask as Fig. 1 is adopted when calculating the power spectra.}
\label{fig:cl}
\end{figure}

\begin{figure*}[!htb]
\centering
\includegraphics[width=1.0\textwidth]{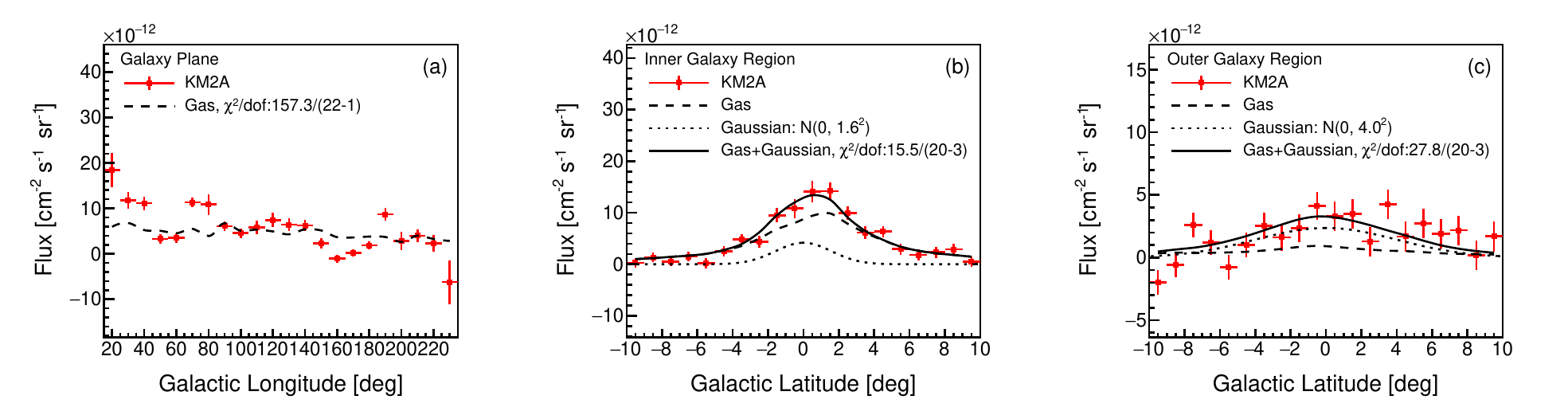}
\includegraphics[width=1.0\textwidth]{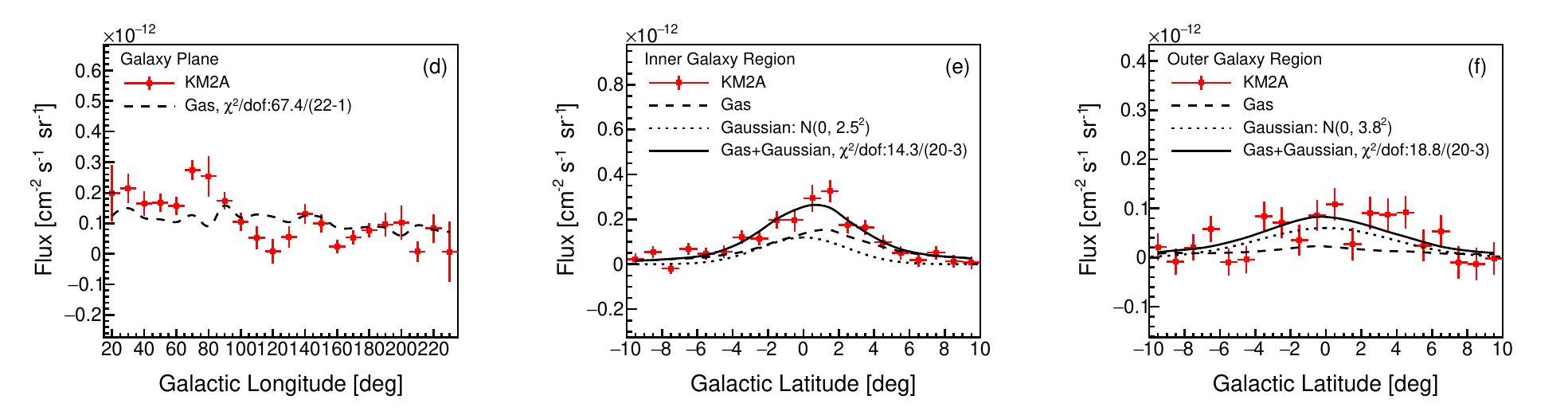}
\caption{Galactic longitude and latitude profiles of the diffuse emission for 
energy bands $10-63$ TeV (top three panels) and $63-1000$ TeV (bottom three panels), 
respectively. When fitting the diffuse emission to a combination of the PLANCK-derived
gas template and a Gaussian distribution along Galactic latitude, the dashed, dotted, and 
solid lines in panels (b), (c), (e), and (f) are contributions from the PLANCK-derived 
gas template, the Gaussian distribution, and their sum, respectively. For panels
(a) and (d), only the gas-related component is shown. The latitude distributions
are not substantially different from those when attributing all the diffuse emission 
to the PLANCK-derived gas template, as in the analysis presented in the main text.}
\label{fig:1d_prof_gauss}
\end{figure*}

The angular power spectrum can be used to statistically compare the distributions of
different skymaps. The results of the angular power spectra of the relative $\gamma$-ray 
flux map ($E>25$ TeV; black line), $(F-\bar{F})/\bar{F}$ where $F$ is the flux in each 
pixel and $\bar{F}$ is the average flux in the ROI, and the relative gas distribution 
derived from the Planck dust opacity (purple line), after applying the same mask as used 
in this analysis, are shown in Fig.~\ref{fig:cl}. The gray band represents the $1\sigma$ 
statistical fluctuations of the data. The gas term is normalized to ensure that the 
expected number of counts 
after convolving the LHAASO instrument response and exposure is the same as the data.
Due to the limited data statistics, the fluctuation may affect the calculation of the
angular power spectrum of the gas term. We thus sample $10^3$ times according to Poisson 
distributions and derive the median value as represented by the purple line. 
It is shown that for $l>10$ the current data give consistent angular power spectrum
with the expectation based on the gas model. At smaller $l$, the data give slightly
higher power than the gas model, which indicates that the data are more clumpy than
the gas distribution. This is in general agreement with the one-dimensional longitude
distribution (Fig. 3). However, the differences under the current statistics are small. 
With the accumulation of data, further exploration along this regard is necessary.

\subsection{Longitude and latitude distributions for alternative fittings}

The PLANCK gas distribution is employed as template to fit the longitude 
and latitude distributions in the main text. To study possible contributions
to the diffuse emission from components other than that proportional to the
gas column density, we add a Gaussian template along Galactic latitudes to 
the fittings. Fig.~\ref{fig:1d_prof_gauss} shows the longitude and latitude 
distributions overplotted with the fitting results. The inclusion of a Gaussian 
template does improve the fitting slightly, compared to attributing all the 
diffuse emission to the PLANCK-derived gas template. However, the difference 
is not significant enough to claim the detection of a new component with 
morphology different from the gas-related component.

\subsection{Local CR spectral fittings}

\begin{figure*}[!htb]
\centering
\includegraphics[width=0.48\textwidth]{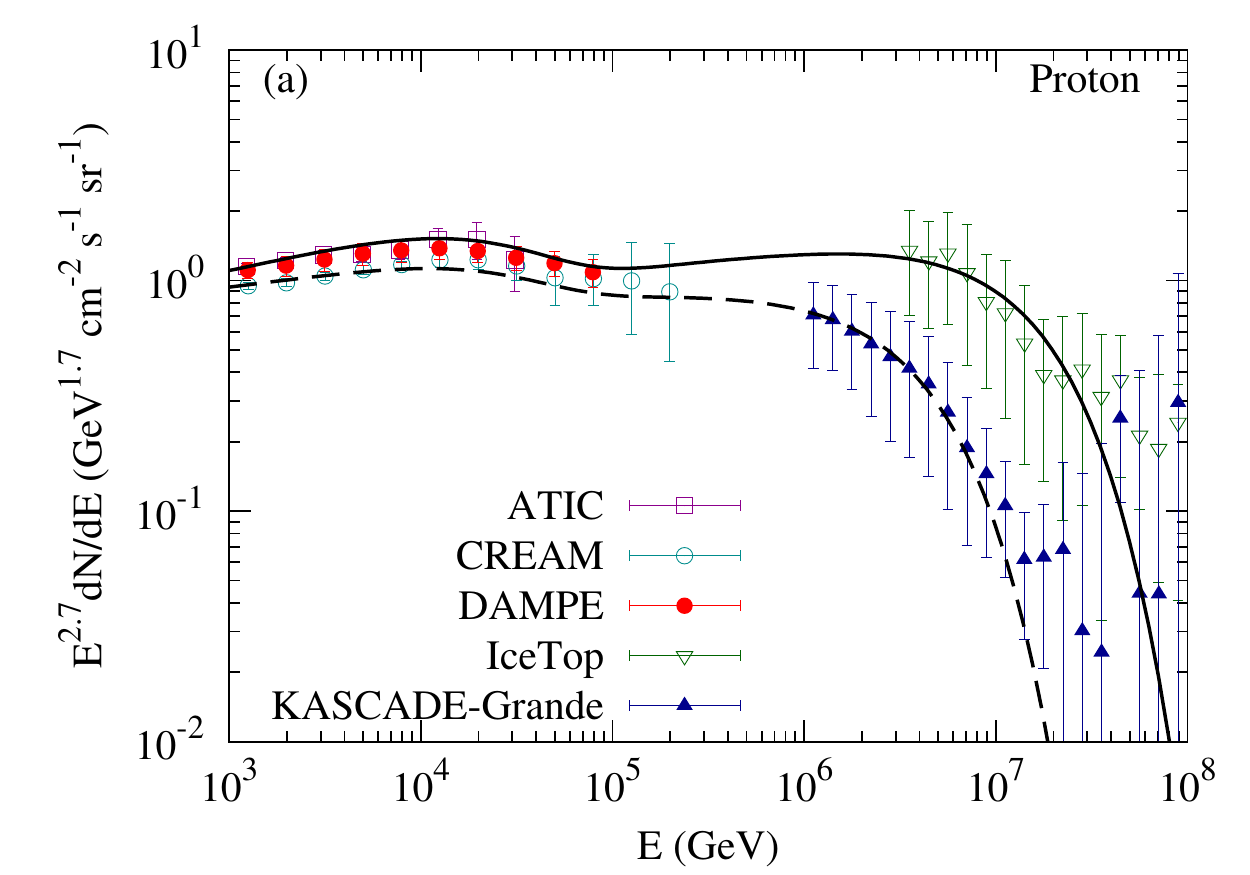}
\includegraphics[width=0.48\textwidth]{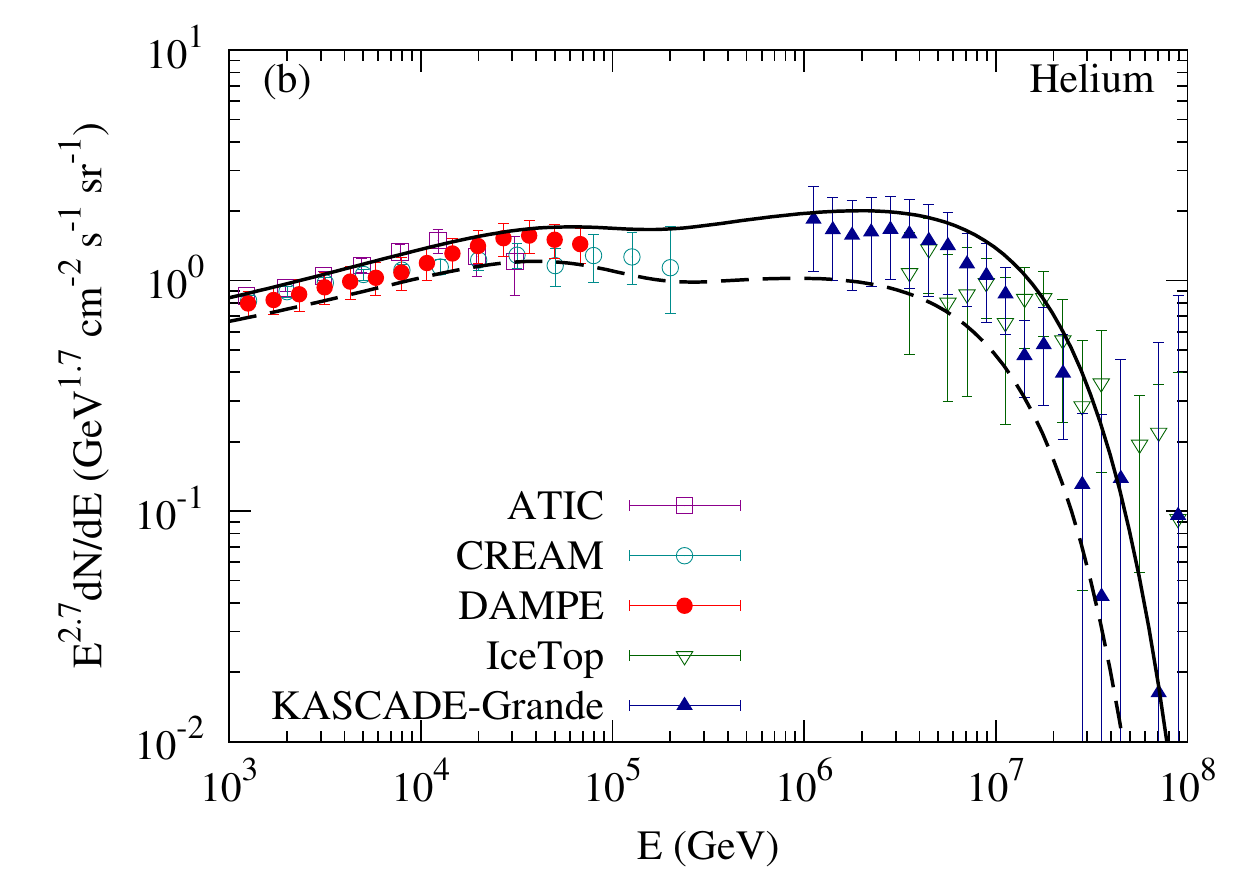}
\caption{Spectra of locally measured protons (a) and helium (b) together with 
phenomenological fittings to represent the uncertainties.}
\label{fig:phe}
\end{figure*}

Fig.~\ref{fig:phe} shows the proton and helium spectra measured locally
by balloon or space experiments [21, 22, 43, 44] and ground-based air 
shower experiments [45, 46]. We use the sum of two exponentially cutoff 
power-law functions 
$$A_1(E/{\rm TeV})^{-B_1}\exp(-E/C_1)+A_2(E/{\rm TeV})^{-B_2}\exp(-E/C_2),$$ 
to fit the data. The parameters (Table~\ref{table:cr}) are adjusted to 
properly reflect the uncertainties of the measurements. The resulting high 
and low representations of the CR spectra are shown by the solid and dashed 
lines, respectively.

\begin{table*}[!htb]
\centering
\caption{Fitting parameters of the local CR spectra.}
\begin{tabular}{ccccccc}
\hline
 & $A_1$ & $B_1$ & $C_1$ & $A_2$ & $B_2$ & $C_2$ \\ 
 & GeV$^{-1}$cm$^{-2}$s$^{-1}$sr$^{-1}$&  & TeV & GeV$^{-1}$cm$^{-2}$s$^{-1}$sr$^{-1}$&  & PeV \\
\hline
Proton-High & $3.40\times10^{-9}$ & $2.35$ & $25.0$ & $5.51\times10^{-9}$ & $2.60$ & $15.0$ \\ 
Proton-Low  & $1.80\times10^{-9}$ & $2.35$ & $25.0$ & $5.71\times10^{-9}$ & $2.66$ & $4.0$ \\ 
Helium-High & $0.75\times10^{-9}$ & $2.10$ & $50.0$ & $5.95\times10^{-9}$ & $2.55$ & $13.0$ \\ 
Helium-Low  & $0.75\times10^{-9}$ & $2.10$ & $50.0$ & $4.55\times10^{-9}$ & $2.60$ & $9.0$ \\ 
\hline
\end{tabular}
\label{table:cr}
\end{table*}

\end{document}